\documentclass{article}
\usepackage{graphicx}
\usepackage{amsmath}
\usepackage{amsthm} 
\usepackage{amssymb}
\usepackage{color}
\usepackage{textcomp}
\usepackage{rotating}
\begin{document}

\title{Scientific requirements of the VAO SED tools}
\author{R. D'Abrusco, J. McDowell and the VAO SED team}

\maketitle

\section{Introduction}
\label{introduction}

\begin{figure}
   \centering
   \includegraphics[width=3in,height=8in]{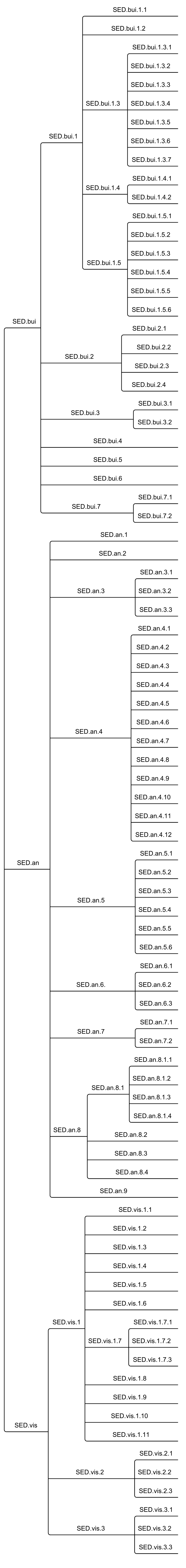} 
   \caption{Tree-graph of the hierarchical structure of the SED tool requirements.}
   \label{plot:sed_requirements_tree}
\end{figure}

This document describes the scientific requirements for the the VAO SED builder and analysis tool. It is divided in three sections describing the 'SED building' (section \ref{sedbuilder},  
'SED analysis' (section \ref{sedanalysis}) and 'SED visualization' (section \ref{sedvisualizationandediting}) capabilities of the tool respectively. Each section is split
in multiple subsections, each one describing a distinct requirement of the tool. Each distinct requirement discussed in this document is associated to a 
label indicating the general section of the document (SED builder - SED analysis - SED visualization) where the requirement can be found, and a unique 
index (for example, {\bf SED.an.3.1} for the first sub-requirement of the third
requirement of the analysis section). These labels are used to provide a quick reference to the different parts of the requirements and provide a handle to the hierarchical 
structure of the document. The hierarchy of requirements is also shown in the tree-graph in figure \ref{plot:sed_requirements_tree}, which is associated to the break-down 
scheme adopted throughout this document. The labels (in boldface in the document) are also used in tables \ref{table:comparison}, \ref{table:prioritizationSEDbui}, 
\ref{table:prioritizationSEDan} and \ref{table:prioritizationSEDvis}.

\section{SED builder}
\label{sedbuilder}

The overall goal of this section of the document is to outline the basic capabilities of the tool regarding the ability to read different data types, their conversion to VO formats
and their combination to create the SED. The Spectral Energy Distribution (SED) of  a source can be built by combining photometric points and spectroscopic segments; the basic definitions of these 
two different type of data given below:
\begin{enumerate}
\item Photometric points. A photometric point is specified, at a very basic but general level, by assigning three numbers $(s, f(s), t)$, namely a spectral coordinate $s$ (either a wavelength, 
a frequency or an energy), a flux $f(s)$ (or flux density, or luminosity) measured at that spectral coordinate, and the time $t$ of the observation. While the time coordinate
associated to a photometric measurement is a fundamental information of its own, in the following the explicit dependence will be dropped for the sake of the simplicity. At the 
same time, the time $t$ of the observation will be considered part of the metadata accompanying every measurement. In the ideal case, error estimates for these values are also given. 
Sometimes, upper limits on $f_u(s)$ are the only available data; in these cases, the points need to be labelled as such but otherwise handled as if they were detected values for this 
section (builder tool) of the software;
\item Spectroscopic segments. A spectroscopic segment consists of a relatively tightly spaced collection of adjacent spectral coordinates and corresponding fluxes (or luminosities): $(s_i, f(s_i))$. 
Even in this case, a time coordinate associated to the measurement of the spectrum is required information.
\end{enumerate}
These data usually can be contained in files in a variety of formats and can be expressed in different unit of measure. It is worth stressing that the apparently simple 
scenario describing the construction of an aggregate SED from distinct elementary data elements (photometric points and/or spectral segments) is complicated, in the real 
world, by multiple effects introduced (mostly) by the instrumental settings of the different observations and, to a lesser degree, by the properties of the emission mechanism 
of the observed source. For example, multiple observations of a single photometric point (i.e., flux) associated to a given spectral coordinate (be it the efficient spectral coordinate 
$s_{{\text{eff}}}$ or a generic spectral interval $[s_1, s_2]$) of the (nominally) same region of a source with (nominal) same apertures and instrumental configurations can disagree for 
multiple reasons:
\begin{itemize}
\item Instrumental effects, that can be split in three different contributions:
	\begin{itemize}
	\item Photometric system effects: small differences in the filters definitions, observation techniques, reduction procedures and absolute calibrations among different observations 
	can introduce large differences in final data, even if this data have been obtained apparently using one single photometric system and the same aperture;
	\item Aperture effects: even slight differences in the fraction of the area of extended sources like galaxies (and in the position of the observed area relative to the source) can lead to 
	significant fluctuations of the values of the observed integrated fluxes;
	\item Crossmatching effects: differences in photometric measurements of a same source with ideally identical instrumental configuration can also arise from inaccurate 
	characterization of the observed positions, leading to incorrect identifications;
	\end{itemize}
\item Intrinsic effects: many types of sources may show intrinsic variability in their emission in some spectral intervals or in the whole SED, leading to scatter in the
observed fluxes measurements even if all the others possible instrumental sources of scatter have been accurately checked and corrected. 
\end{itemize}
In order to take into account all these possible effects, any available metadata (time, filter definition, astrometry of the observation, instrumental configuration, spatial model of the source, 
reduction parameters, etc.) associated to each of the observed photometric points and spectral segments is valuable information that should be accessible to the user during any phase of the 
process of construction of the SED.

\subsection{Access to data}
\label{accesstodata}

A key element of the characterization of the data elements used in the construction of the aggregate SED is the origin of data. In general, two categories 
of data can be distinguished in terms of the their origin and the reduction/analysis steps performed before they are ingested by the SED builder tool:
\begin{itemize}
\item {\bf Data owned by the user}: the user has observed his own data (images, spectra, spectral cubes, time series) that are locally stored (on his computer). 
These data have been reduced using user's own pipeline, with observational parameters estimated using the user's analysis package and set-up and stored 
as tabular data in files in a generic tabular format; 
\item {\bf Data elements accessible through VO protocols}: images (SIAP), spectra and spectral data cubes (SSAP), spectral lines (SLAP) and associated 
(or independent) photometric points and spectra encoded as tabular data (TAP), retrieved through any of the services offering an implementation of the fundamental VO query protocols:
	\begin{itemize}
	\item Simple Image Access (SIAP): {\it...a protocol for retrieving image data from a variety of astronomical image repositories through a uniform interface. The interface is 
	meant to be reasonably simple to implement by service providers. A query defining a rectangular region on the sky is used to query for candidate images. The 
	service returns a list of candidate images formatted as a VOTable...}\footnote{Excerpt from IVOA website at  the URL http://www.ivoa.net/Documents/SIA/};
	\item Simple Spectral Access (SSAP): {\it...a uniform interface to remotely discover and access one dimensional spectra. SSA is a member of an integrated family of data
	access interfaces altogether comprising the Data Access Layer (DAL) of the IVOA. SSA is based on a more general data model capable of describing most tabular
	spectrophotometric data, including time series and spectral energy distributions (SEDs) as well as 1-D spectra...}\footnote{Excerpt from IVOA website at the URL 
	http://www.ivoa.net/Documents/latest/SSA.html};
	\item Spectral Line Access (SLAP): {\it...a protocol for retrieving spectral lines coming from various Spectral Line Data Collections through a uniform interface within the 
	VO framework. These lines can be either observed or theoretical and will be typically used to identify emission or absorption features in astronomical spectra}\footnote{Excerpt from
	the IVOA website at the URL: http://www.ivoa.net/Documents/latest/SLAP.html}
	\item Table Access (TAP): {\it a service protocol for accessing general table data, including astronomical catalogs as well as general database tables. Access is provided 
	for both database and table metadata as well as for actual table data...}\footnote{Excerpt from IVOA website at the URL http://www.ivoa.net/Documents/TAP/}.
	\end{itemize}
Usually, such publicly available data have been reduced and/or analyzed with their own custom pipelines, softwares and parameter configurations that not necessarily 
are agreed upon by all the users for all possible research scenarios and projects. On the other hand, some of these data, for example spectral data, can be simulated 
and are not characterizable in terms of the reduction process. 
\item {\bf Precomputed SEDs}: photometric and spectral information covering all the EM spectrum or a large spectral interval can be provided by large scientific
collaborations (like CANDELS\footnote{Official website at the URL http://csmct.ucolick.org/}) or by archival services (like NED\footnote{More details at the URL 
http://nedwww.ipac.caltech.edu/forms/photo.html}) at different levels of refinements of the raw data:
	\begin{itemize}
	\item `Survey' SEDs: statistically consistent representations of the SEDs obtained by collecting and federating the original data observed from a unique survey or a collection 
	of carefully combined complementary surveys (compare with the rebinned SEDs defined in \ref{convertanaggregatesed});
	\item  `Collection' SEDs:  representations of the SEDs obtained by collecting and federating the original data taken from different inhomogeneous observations with no or little
	effort in the process of blending the different features of the data elements (compare with the aggregate SEDs defined in \ref{assembleahetero}).
	\end{itemize} 
\end{itemize}
The SED builder tool shall be required to provide the user with the access to data of all these different types in its final version. More specifically, the SED tool requirements for data reading can 
be specified as follows: 
\begin{itemize}
\item The SED tool shall be able to ingest local data ({\bf{SED.bui.1}}) owned by the user for the specific data types described in the paragraphs \ref{convertphotometry} ({\bf{SED.bui.1.1}}), \ref{convertspectra} ({\bf{SED.bui.1.2}}), 
\ref{extractaspectrum} ({\bf{SED.bui.1.3}} and related subtasks), \ref{convertatheoretical} ({\bf{SED.bui.1.4}} and associated sub-requirements, and \ref{extractaphotometryfromanimage}({\bf{SED.bui.1.5}}). Since it is possible that 
part of the data owned by the user is not stored in VO format files, the SED builder tool shall be required to be capable of ingesting data in files of both VO-compliant formats and Comma Separated Values (CSV) and 
Tab Separated Values (TSV) formats; 
\item The SED tool shall be able to provide standardized access to the archival data elements ({\bf{SED.bui.2}}) exposed through services implementing the specifications of the IVOA protocols for data 
retrieval described above\footnote{An example of the type of data access interfaces required and already implemented in other tools can be found at the following URL: 
http://www.star.bris.ac.uk/$\sim$mbt/topcat/sun253/sun253.html$\#$vo-windows}. Such data elements will be retrieved using different search criteria (for both single source and multiple sources at once):
	\begin{itemize}
	\item Images (SIAP) ({\bf{SED.bui.2.1}}): name of the source(s) or sky coordinates (example: all images available for the source ``3C273" or for the coordinates (ra=152.000042, dec=7.504556));
	\item Spectral data (SSAP) ({\bf{SED.bui.2.2}}): name of the source(s) or sky coordinates (example: all spectral data available for the source ``3C273" or for the coordinates (ra=152.000042, dec=7.504556));
	\item Spectral line data (SLAP) ({\bf SED.bui.2.3}): wavelength range where the spectral lines can be found in rest frame (example: all spectral lines in the spectral interval $[4300, 400] \AA$);
	\item Tabular data (TAP) ({\bf{SED.bui.2.4}}): name of the source(s), sky coordinates and specific constraints relative to the type of data retrieved (for example: all fluxes available for the sources ``3C273" or
	measured for sources detected around the coordinates (ra=152.000042, dec=7.504556) within a radius of 0.5', or all fluxes measured for all the available sources with spectroscopic 
	redshifts $z_{spec} \in [0.15, 0.16]$ and flux in the XMM broad band higher than $11.5\cdot 10^{-11} {\text {erg}}/({\text {cm}}^2\cdot {\text s})$ etc. etc.);
	\end{itemize}
\item The SED tool shall be required to support the access to precomputed SEDs ({\bf{SED.bui.3}}), provided as  ``Survey" products by specific scientific collaborations, like in the CANDELS case, for 
extragalactic sources ({\bf{SED.bui.3.1}}), or as ``Collection" SEDs provided by external services using archival data like the NED SEDs service  ({\bf{SED.bui.3.2}}). Even in this case, the search interface 
to this datasets shall provide multiple search criteria: name of the source, sky coordinates of the sources and generic constraints on the observational parameters associated to the source. 
\end{itemize}
Since most potential users are adamant that they do not entirely trust the data reduction and analysis done ``elsewhere" by ``anyone else" at least for the type of data 
that they are used to work with, the ability to read in the data owned by the user, together with the support of the access to VO-published data,  
is a high priority in order to offer as soon as possible limited but fully working tool to the astronomical community.

\subsection{Interoperability with VO tools}
\label{interoperability}

A fundamental requirement of all tools and services developed by the VAO is the ability to interoperate and communicate seamlessly. The {\it de facto} standard protocol 
for interoperability among astronomical tools is the Simple Application Messaging Protocol (SAMP)\footnote{More details can be found at the URL webpage http://www.ivoa.net/Documents/latest/SAMP.html}, 
and developed on behalf of the IVOA. Most of the tools for the analysis, visualization and retrieval of astronomical data have already implemented the SAMP protocol.
For this reason, a requirement of the SED tool shall be the support of the SAMP protocol ({\bf{SED.bui.4}}), in order to let the SED building and analysis tools to get along with 
external applications. Some of these applications can be used to accomplish some of the requirements described in the following section of this document. A list of such 
softwares, with a very short description of their main capabilities, can be found in the following list:
\begin{itemize}
\item TOPCAT\footnote{Official website at the URL http://www.star.bris.ac.uk/$\sim$mbt/topcat/}: the Tool for OPerations on Catalogues And Tables is an interactive graphical tool for the manipulation and 
visualization of tabular data (this tool could be used specifically to address entirely or in part some of the requirements of this document, i.e {\bf SED.bui.1.1}, {\bf SED.vis.3.1}, {\bf SED.vis.3.2});
\item DS9\footnote{Official website at the URL http://hea-www.harvard.edu/RD/ds9/}: application for the visualization of astronomical images (as above, this tool could be used specifically 
to address entirely or in part some of the requirements described in this document: {\bf SED.bui.1.5.2}, {\bf SED.vis.3.3});
\item Aladin\footnote{Official website at the URL http://aladin.u-strasbg.fr/}: an interactive software for the visualization of digitized astronomical images ({\bf SED.bui.1.5} thanks to its integration with 
SExtractor, {\bf SED.vis.3.1}, {\bf SED.vis.3.1}, {\bf SED.vis.3.2}, {\bf SED.vis.3.3});
\end{itemize}
Other tools which could be useful for the SED tool are not yet SAMP-ified, but are planning to add interoperability capability in the near future. 

\subsection{Usage metrics}
\label{usagestatistics}

The SED tool shall provide logs of usage metrics ({\bf{SED.bui.5}}) in support of the Projects Metrics requirement as described in Section 1 of the
VAO Project Execution Plan (page 40): {\it The VAO will record metrics that measure the growth in scientific usage of its services, 
responsiveness to users, and quality of its services. The metrics will be reported to NSF and NASA as part of the Quarterly and 
Annual Reports.}

\subsection{Conversion of photometric measurements in a table to a VO-compliant format}
\label{convertphotometry}

This tool shall be required to convert a table ({\bf{SED.bui.1.1}}) stored in a file in one of the typical tabular data file format (ASCII, CSV, FITS, XLS) and containing 
photometric data (basically the $(s_i, f(s_i))$ columns) to a VO-compliant format (VOtable or FITS of the particular forms  described in the IVOA Spectral Data Model).
The tool shall also be able to recognize the presence of metadata in the header of the input file and transfer them to the header of the VO-format file created 
as output. An example of a program which can perform similar operations (with a different scope and with limitations) is the {\it tcopy} command contained 
in STILTS, a set of command-line tools for general table manipulation based on the STIL library.\footnote{Command description at the URL: http://www.star.bris.ac.uk/$\sim$mbt/stilts/sun256/tcopy.html}.

\subsection{Conversion of spectra in various observational formats to a VO-compliant format}
\label{convertspectra}

The tool shall also be required to read in a file containing spectroscopic data ({\bf{SED.bui.1.2}}) in one of the most common formats found in the literature (again ASCII, 
CSV, FITS, XLS) and convert the data to one of the VO-compliant formats, i.e. either VOTable or FITS format. This capability is very similar to the 
previous one, given that the nature of the tabular data does not change for photometry or spectroscopy.  The supported input formats will include the common 
IRAF flavors of FITS in which the spectrum, errors and quality flags are encoded in a single 2D primary array (in which the first axis is wavelength and the second 
axis is ``kind of thing".)

\subsection{Extraction of a spectrum from spectral data-cube}
\label{extractaspectrum}

The SED builder tool shall be able to handle complex data in the form of spectral data cube ({\bf{SED.bui.1.3}}), i.e. not simple two dimensional tables containing $(s_i, f(s_i))$ columns representing spectral coordinates 
and the corresponding measures of flux (as in the cases described in the first two points) but at least three-dimensional data, generally $(s_i, f(s_i), x_i)$ where $x_i$ 
is either a scalar or a two components vector associated to an additional spatial or time variable. Multiple different cases can be distinguished:

\begin{itemize}
\item Slit spectroscopy ({\bf{SED.bui.1.3.1}}): the additional coordinate is a scalar and represent a spatial coordinate (for example, in slit spectroscopy, a continuum of 
spectra with the dispersion axis perpendicular to the direction of the slit is produced, so that the spectra can be seen as a function of the position 
along the slit). In this case, the spectral cube needs to express the spatial resolution along the axis of the slit, so that each point is represented by 
three dimensional vectors $(s_i, f(s_i), x_i)$. This tool is required to be able to extract the spectrum $(s^*_i, f(s^*_i))$ associated to a given position 
$x^*$ along the slit, i.e. $(s^*_i, f(s^*_i)) = (s_i, f(s_i), x^*)$; 
\item Slitless spectroscopy ({\bf{SED.bui.1.3.2}}): the extraction of spectra from slitless spectroscopy data (for example, grism images) is a multi-stage process 
involving different reduction steps, and for this reason, significantly more complex than the other kind of spectroscopic data. The fundamental problem 
is that there is no a-priori geometrical information about the sources generating the spectral traces seen in the grism image; a direct reference image 
(or a catalogue of sources extracted from the same field) is necessary to determine the reference origins of the spectra, associate the spectroscopic traces
to the sources position and extract the corresponding 2-dimensional spectra from the original grism image. The ability to extract spectra from slitless spectroscopy
data is considered a requirement for the tool and, if possible, it should be referred to a specific tool for spectral data reduction, which should 
also be usable for the reduction of other complex spectroscopic data (for example, echelle spectroscopy data);
\item Integral Field Spectroscopy (IFS) ({\bf{SED.bui.1.3.3}}):  these data are typically produced by instruments with combined spectroscopic and photometric capabilities, 
which produce two dimensional spatially resolved spectra of a region of a source. Spectra are observed in a two dimensional field and stored in a 
4-dimensional table where points are represented by vectors $(s_i, f(s_i), (x_i, y_i))$, where the two spatial coordinates, usually regularly
spaced, determine the central points of the regions emitting the corresponding spectrum $(s_i, f(s_i))$. In order to extract correctly the spectrum at a 
given spatial point, the knowledge of the aperture of the instrument is required. In addition, another requirement to the tool is that it will be capable 
of supporting the extraction of two different types of spectra as the user select a finite region of the image (circular) $C$ or a single ''pixel'' or ''point'' of the 
image. In the former case, the tool will return an averaged spectrum from the region selected $((s_i, \left\langle{f(s_i)}\right\rangle, (x_i, y_i))$ with $(x_i, y_i)$ in $C$), 
while in the latter it will return the raw spectrum associated with the pixel selected, i.e. $((s_i, f(s_i), (x^*, y^*))$. In addition to these simple 
geometrical constraints, the tool shall support the correction of the extracted spectra on the basis of a theoretical or empirical spatial distribution $F^{(the,emp)}(x, y)$ 
or $F^{(the,emp)}(r)$ associated to the underlying extended source (in a given spatial coordinate system), so that the spectra observed in different positions can 
be calibrated for the luminosity profile:
\begin{eqnarray}
(s_i, f(s_i), (x_i, y_i)) \rightarrow (s_i, f^{(corr)}(s_i), (x_i, y_i)) \text{    given    } F^{(the,emp)}(x, y)\\
(s_i, f(s_i), (r_i)) \rightarrow (s_i, f^{(corr)}(s_i), (r_i)) \text{    given    } F^{(the,emp)}(r)
\end{eqnarray}
\item Fibers spectroscopy ({\bf{SED.bui.1.3.4}}): in this case, spectra are observed for $N$ positions $(x_i, y_i)$ in a given field, and the light coming from these 
regions is transferred from the focal plane of the telescope to the spectrograph through $N$ distinct optical fibers. Similarly at what happens for the IFS, 
such spectra can be stored in a 4-dimensional table where points are represented by vectors $(s_i, f(s_i), (x_i, y_i))$. In this case, spectra can be either 
unrelated (when multiple distinct sources are observed in the same field at the same time), or belonging to the same extended source observed in 
different points. A difference with the previous case is that, for fibers spectroscopy, the choice of the positions of the fibers is completely free, i.e. not regularly 
spaced as in the IFS case, since the uninteresting regions of the field are masked;
\item `Photometry-based' spectroscopy ({\bf{SED.bui.1.3.5}}): some type of observations, obtained in peculiar spectral ranges of the EM spectrum and with suitable 
instrumental configuration, provide detailed knowledge of the main parameters of each photon reaching collected by the optical section of the instrument and measured 
by the detector. For example, in X-rays astronomy, each photon is associated to a unique arrival position on the detector, energy and arrival time, so that a collection 
of events or photons can be represented as peculiar spectral data cube $P$, i.e. a four dimensional array, where each element $p_{i}$ corresponds to a vector:
\begin{equation}
P \rightarrow p_i = p_i(\mathbf{x}_i, E_i, t_i)
\end{equation}
In such case, the difference between photometry and spectroscopy for a given sources becomes blurry, since both types of data (images and spectra) can 
be extracted from the spectral data cube. In general, the image $I_{S,T,B}(\mathbf{x})$ of a region of a source $S$, with a given integration time $T$ and inside a given energy band $B$, is 
obtained by integrating the subset of the 'spectral data cube'  corresponding to the spatial region $S$ along the time and energy axes, while the spectrum $f(E)_{S,T}$from $S$ is
the result of the integration only along the time coordinate:
\begin{eqnarray}
I_{S,T,B}(\mathbf{x}) = \int_{T}\int_{B} (P\cap S)dtdE\\
f_{S,T}(E) = \int_{T}\int_{S} P dtd\mathbf{x}
\end{eqnarray}
\item Time dependent spectroscopy ({\bf{SED.bui.1.3.6}}): for a given spectral configuration (i.e., slit spectroscopy, slitless spectroscopy, fiber spectroscopy, etc.), the 
additional parameter is a time coordinate, and the spectral cube contains multiple spectra $(s_i, f(s_i))$ observed at different times $t_i$ for each of the.  
The tool shall be required to extract from the spectral data cube the single spectrum $(s^*_i, f(s^*_i))$ associated to the time $t^*$, i.e. $(s^*_i, f(s^*_i)) = (s_i, f(s_i), t^*)$.
\end{itemize}
\noindent The SED tool shall be required to handle all the spectra data cubes described above. 
The detailed specification for \ref{convertphotometry} to \ref{extractaspectrum} above is implicit in the IVOA Spectral Data Model, defining the output, and a set of representative 
input files which we will have to collect. 

\subsubsection{Extraction of photometric points from a spectral data cube}
\label{extractaphotometry}

This is a special case of the above function, with the additional functionality that the spectrum is convolved with a user-supplied photometric filter ({\bf{SED.bui.1.3.7}}).

\subsection{Conversion of a theoretical spectral model to a VO compatible spectrum}
\label{convertatheoretical}

The SED builder tool shall be required to be able to read in a theoretical spectral model ({\bf{SED.bui.1.4}}) (for example, a collection of basic information containing the source model, the best-fit parameters 
of the model and the spectral coordinate range the model applies to, the method used to derive the estimates and all the applicable metadata) and produce a VO-compliant file 
containing a tabular representation of the spectrum generated by the model ({\bf{SED.bui.1.4.1}}) according to the following scheme: 
\begin{equation}
(\text{model}(s), \text{range}(s), \text{method}, \vec{p},...) \rightarrow (s_i, f(s_i)) \rightarrow \text{VOtable, FITS}
\end{equation}
\noindent where $\vec{p} = (p_1, p_2,..., p_M)$ is a vector containing the values of the $M$ parameters of the model. 
In order to implement this functionality, the following parameters need to be determined:
\begin{itemize}
\item the format for the functional form of the source model (candidate: Sherpa model definition language);
\item the format and allowed data types for the parameter names and values;
\item a method for supplying metadata parameters (the possible parameters are described in the VO Spectral Model document, although some will not be relevant for theoretical data and we should enumerate those);
\item the format for describing the spectral coordinate grid;
\item the nature and format of any additional info specific to theoretical models.
\end{itemize}
\noindent Another requirement for this tool shall be the ability of producing, given a single theoretical spectral model, multiple representations of the spectrum generated by the 
model ({\bf{SED.bui.1.4.2}}), by evaluating the same model on a grid generated by one (or more) of the parameters of the model. Given two parameters $p_1$ and $p_2$, for 
example, the model should be evaluated on the grid of values:
\begin{multline}
\{[p_1^{(1)}, p_1^{(2)},..., p_1^{(K)}] \times [p_2^{(1)}, p_2^{(2)},..., p_1^{(L)}]\} = \\
[(p_1^{(1)}, p_2^{(1)}), (p_1^{(1)}, p_2^{(2)}),...,(p_1^{(2)}, p_2^{(1)}),(p_1^{(2)}, p_2^{(2)}),...,(p_1^{(K)}, p_2^{(L)})]
\end{multline}
The product will be a set of $K\times L$ distinct realizations of the model:
\begin{multline}
(\text{model}(s), \text{range}(s), \text{method}, \vec{p},...) \rightarrow \\
\{(s_i, f(s_i, (p_1^{(1)}, p_2^{(1)}))), (s_i, f(s_i, (p_1^{(1)}, p_2^{(2)}))),...,(s_i, f(s_i, (p_1^{(K)}, p_2^{(L)})))\} \rightarrow \text{VOtable, FITS} 
\end{multline}
The set of realizations of the model can be represented as a spectral data cube, but since no official spectral data cube data model from the IVOA is available at this time, we could 
require the tool to just be able to save all the distinct realizations of the model spectrum in a single VO-compliant file. The tool shall also be 
required to be able to save the distinct realizations of the spectral model using a not-standard definition of data-cube model (waiting for an official specification from IVOA). 
In such case, if only one parameter $p_1$ of the model is let varying on the grid,  the 'data cube' will be three-dimensional and similar to the spectral 
data cubes produced either by slit spectroscopy or time-dependent spectroscopy (see paragraph \ref{extractaspectrum}). If the grid is 2-dimensional,  
then the spectral data cube will be 4-dimensional and similar to the spectral data cube produced by IFS (two spatial coordinates $(x_i, y_i)$). 
The SED tool, at this later stage and hopefully with the emergence of a standard spectral data cube model definition, shall be able to save the spectral data 
cube associated to the different realizations of the model in a VO-compliant format.

\subsection{Extraction of a photometric point from an image}
\label{extractaphotometryfromanimage}

The SED builder tool shall be required to be able to extract photometric parameters from an image and save the resulting parameters in a VO-compliant format ({\bf{SED.bui.1.5}}). 
A two dimensional image with pixel values $n(x_i, y_i)$ can be associated to each of two main sub-cases: 
\begin{enumerate}
\item the Poisson case, where the image pixel values are in instrumental (count) units with a user-supplied conversion from counts to flux. In this case, 
the confidence intervals (at a given level of confidence) are determined using Poisson statistics;
\item the Gaussian case, where image pixel values are already in flux units (e.g. FITS BUNIT keyword specified). In this case confidence intervals (at a 
given level of confidence) can only be determined if a separate error image is also supplied.
\end{enumerate}
The SED builder tool shall be required to support the extraction of photometric parameters in two different situations:
\begin{itemize}
\item The main input is an image array, possibly with associated background array and
error array of the same dimensions ({\bf{SED.bui.1.5.1}});
\item Use an externally supplied region description S ({\bf{SED.bui.1.5.2}}). The flux will be integrated over
the pixels in this region. (A simple case is a circular region specified as a celestial position and a radius). This 
region may have been created manually by the user or be part of the output of a source detection program such 
as Sextractor\footnote{Official webpage at the URL http://www.astromatic.net/software/sextractor}.
\end{itemize}
\noindent Other requirements of this task of the SED tool are the following:
\begin{itemize}
\item Optionally, accept a second region description to define a background region B ({\bf{SED.bui.1.5.3}});
\item Determine the value of the flux (background-subtracted if background is supplied) and the associated uncertainty from a region B of an image provided to the tool ({\bf{SED.bui.1.5.4}});
\item Apply an aperture correction (a scalar value supplied by the user) to the flux value determined from the region B in the image ({\bf{SED.bui.1.5.5}});
\item Read the relevant metadata from the input image to construct an IVOA Spectrum instance (with one data point) with the appropriate metadata entries ({\bf{SED.bui.1.5.6}}) and write out 
the IVOA Spectrum instance using a VO-compliant format;
\end{itemize}

\subsection{Assembling a heterogeneous SED dataset - SED aggregate from individual photometry and spectral segments}
\label{assembleahetero}

A fundamental requirement of this tool will be the ability to assemble an aggregate SED ({\bf{SED.bui.6}}) from distinct data elements (whatever their origin), consisting in separate 
photometric points $(s_i, f(s_i))$, and spectral segments. The functionality to convert  data from its original form to VO compliant Spectrum segments is assumed to be provided by 
the requirements described in the paragraphs \ref{extractaspectrum} to \ref{extractaphotometry}. The functionality covered here is to aggregate the individual segments in a single 
file, possibly combining information as specified by a future VO SED standard. The single elements are required to satisfy the following criteria:

\begin{itemize}
\item The input segments must already all have the same observable y-axis quantity (flux, luminosity, surface brightness) since
there is no generic way to interconvert those;
\item The input segments may have different units in either spectral  coordinate or flux;
\item The input segments may have different x-axis quantities (wavelength, frequency, etc.);
\item The aggregator will have an option to perform unit conversions so that all the segments end up with the same units and 
same x-axis quantity;
\item An aggregated SED with heterogeneous units is also a valid dataset, so the unit conversion option will be able to be 
turned off by the user.
\end{itemize}

\noindent It is important to stress that different segments may overlap in spectral coordinates. Note also that different segments will usually have different 
observation dates, and these must be retained. The simplest version  of this tool will just concatenate, for example, the $\left\langle{\text{TABLE}}\right\rangle$ 
elements of the separate input VOTABLEs (or FITS tables). Note that the VOTABLE  standard allows multiple $\left\langle{\text{TABLE}}\right\rangle$ within 
one $\left\langle{\text{VOTABLE}}\right\rangle$, just like the FITS standard allows multiple $\left\langle{\text{TABLE}}\right\rangle$ within one $\left\langle{\text{FITS}}\right\rangle$. 
Additional functionality includes the unit conversions,  x-axis conversions, and possibly consolidation of metadata between compatible segments. This latter 
function would, for example, combine U, B, V photometry segments which share the same instrument and observing date into a single segment; that process 
will need to be spelled out in detail in an IVOA SED standard. The very nature of the instrumental and source-intrinsic effects involved in different observations 
can change the measured values of the photometric points at the same (at least nominally) spectral interval, thus introducing a scatter in the distribution of such 
values (as summarized in section \ref{sedbuilder}). While the higher priority requirement of the SED builder tool shall be the ability described above to `stick 
together' all the available data elements for a given source in an aggregate SED (assuming that all such data have been properly fluxed and corrected) at a 
lower priority, the tool shall support the creation of a ``fine-tuned" version of the aggregate SED, where the discrepancies between different data points have been 
corrected by invoking some of the analysis capabilities described in the next section \ref{sedanalysis}. 
%
%
%
%

\subsection{Aperture correction on photometric points}
\label{aperturecorrectionphotometric}

The effect of the inclusion of distinct individual photometric points in an aggregate SED depends crucially on the aperture used for the different 
observations. In general, two different cases may be described: for point-like sources, the aperture of the measurement may include a finite fraction of the 
instrumental Point Spread Function (PSF), while in the case of extended sources, the total brightness of the source can be inferred from the fraction of the 
source actually observed. In both cases, the correction depends on the comparison of the aperture of the observation with a spatial model of the light 
emission, be it the instrumental PSF model for the point sources or the assumed spatial emission model of the source for intrinsically extended sources . 
For this reason, the aperture correction does not depend on the photometric filter of the observation but on the individual data points and so, in order to 
handle it correctly, a certain amount of information is required for each single photometric point. For example, in the proposed Photometry Data Model 
specification\footnote{IVOA Photometry Data Model document, version 0.2} by J. McDowell, every photometric point can be accompanied by two optional 
metadata expressing the point source aperture fraction and the aperture fraction correction actually applied, according to the formula:
\begin{equation}
f_{meas} = \mathrm{ApFrac}\cdot f_{Tot}
\label{eq:correction}
\end{equation}
\noindent where the Aperture Fraction ApFrac is between 0.01 and 1.0. This equation holds for point sources, while for extended sources the correction
applied can be different from the point source corrections. Another useful metadata described in this specification of the Photometry DM is a flag indicating
whether an aperture correction has been performed or not. The SED tool will be required to display interactively this type of metadata, if available (among 
the other  metadata, as described in section \ref{sedvisualization}) ({\bf SED.vis.1.7}). The SED tool will also be required to perform a simple multiplicative 
correction to the value of the flux of a photometric point following the equation \ref{eq:correction} letting the user choose the value of the ApFrac parameter 
in the $[0.0, 1.0]$ interval ({\bf SED.bui.7.1}), similarly at what required for photometric parameters extracted from image in section \ref{extractaphotometryfromanimage}
({\bf SED.bui.1.5.5}). With a lower priority, the SED tool will also be required to provide the user with the capability to perform advanced aperture corrections 
by taking into account the assumed underlying spatial model for the observed source ({\bf SED.bui.7.2}). In general, given $M$ measured values of the flux 
of a source obtained in the (nominally) same spectral interval:
\begin{equation}
s^{(real)} \rightarrow \{f^{(1)}, f^{(2)},...,f^{(M)}\}
\end{equation} 
\noindent measured at slightly differing physical positions of the observed source (relative to a given cartesian or radial coordinates system):
\begin{equation}
\{(x^{(1)}, y^{(1)}), (x^{(2)}, y^{(2)}),...,(x^{(M)}, y^{(M)}) \} \text{    or    } \{r^{(1)}, r^{(2)},...,r^{(M)}\}
\end{equation}
\noindent with $M$ different filters $\{(s, B^{(1)}(s)), (s, B^{(2)}(s)),...,(s, V^{(M)}(s))\}$, where $B^{(i)}(s)$ is the transmittance curve of the $i$-th filter, 
and generally different apertures $\{A^{(1)}, A^{(2)},..., A^{(M)}\}$. By letting the user choose a simple or composite spatial model for the emission of the 
source, either theoretical or empirical, defined as $F^{(\mathrm{the},\mathrm{emp})}(x, y)$ or $F^{(\mathrm{the},\mathrm{emp})}(r)$, the SED building tool 
shall be able to evaluate the corrections to each of the measured values of the flux, relative to a chosen ``template" aperture $\tilde{A}$ and filter definition 
$\tilde{B}(s)$, thus producing a set of corrected set of flux values:
\begin{equation}
s^{(real)} \rightarrow \{f^{(1)}, f^{(2)},...,f^{(M)}\} \rightarrow \{f^{(corr, 1)}, f^{(corr, 2)},...,f^{(corr, M)}\}
\end{equation}   
\noindent Such ``calibrated" flux values of the ``fine-tuned" aggregate SED can be used to create the rebinned version of the SED, as discussed in paragraph 
\ref{convertanaggregatesed}.

\section{SED Analysis Tools}
\label{sedanalysis}

The tool is required to produce and perform analysis of two basically different types of SEDs: the aggregate SED and the rebinned
SED. These types differ for the type  of information contained: while the {\bf aggregate SED} is a relatively raw collection of all photometric and 
spectroscopic data (points and segments) available for a given source (see  description of the requirements of the SED tool for the creation of an 
aggregate SED - point \ref{assembleahetero}), the {\bf rebinned SED} is obtained by binning an aggregate SED and provide the flux as a function 
of equally spaced spectral coordinate points on a linear scale (or at least, the flux is to be linearly interpolated between adjacent points). Information is 
lost during the rebinning process, so every step needs to be carefully tracked and recorded in  the file containing the final {\bf rebinned SED}.

\subsection{Read in aggregate or rebinned SED}
\label{readinaggregate}

The SED analysis tool shall be required to read in a file containing an aggregate SED ({\bf{SED.an.1}}) or rebinned SED ({\bf{SED.an.2}}). The aggregate SED can
contain different components (photometric points  and spectral segments) with differing units of measure, while a rebinned SED is already a homogeneous list of points specified 
by spectral coordinates and flux/flux density/luminosities  measurements $(s_i, f(s_i))$. In both cases, the SED can be accompanied by additional information regarding the spatial 
extent of the source, as the SED can be produced by either an unresolved (point-like) source, a resolved (extended) source, all-sky signal or be the result of a theoretical model of a simulation
(for example, the SED may have been extracted from a  spectral data cube in a given region of a field). For this reason, the tool shall be required to be able to read in an aggregate 
({\bf{SED.an.1}}) or rebinned SED ({\bf{SED.an.2}}) and to convert the units of measure of the SED for both spectral and flux coordinates (for example, flux - wavelength or flux - frequency, 
photon flux - wavelength  and photon flux - frequency) ({\bf{SED.an.1.1}}). The SED visualization tool shall also be able to display the SED with the units of measures chosen by the user 
({\bf{SED.vis.1}}) and provide a graphical representation of the spatial model of the source associated to the SED, or of the different apertures of the measurements composing the SED,
if such information are contained in the metadata of the aggregate SED ({\bf{SED.vis.3.3}}).

\subsection{Conversion of an aggregate SED dataset to a rebinned SED}
\label{convertanaggregatesed}

One of the basic requirement of the SED tool is the ability to convert from an aggregate SED to a rebinned SED ({\bf{SED.an.3}}). This can be accomplished after the interpolation of the aggregate SED 
which allows to reconstruct the flux values in equally spaced spectral coordinates:  from a collection of photometric points and spectral segments $(s_i, f(s_i)) \rightarrow (s, i(s))$, where $i(s)$ 
is the interpolating function. The SED tool is also required to offer different techniques for interpolation (linear function, polynomials, spline, etc) and should take carefully into account the different 
statistical properties of the spectro-photometric data in the overlapping regions, if any, of the aggregate SED. More specifically:

\begin{itemize}
\item The input data consists of $m$ segments, with $\text{i} = \{1,..., m\} $, each with some number $n_i$ of data points $j = \{1,..., n_i\}$.
Note that $n_i = 1$ if segment $i$ is a photometric point; $n_i>1$ for a spectrum segment. Each data point includes a spectral coordinate
$x_{ij}$, a flux value $F_{ij}$ and possibly a flux confidence interval $(F1_{ij},F2_{ij})$ to a specified level of confidence (and perhaps
other information as specified in the data model). The goal is to map this to a new spectral coordinate grid of N points
$x_k$ where $k \in \{1,..., N\}$ with corresponding fluxes $F(x_k)$ and confidence intervals $(F1(x_k),F2(x_k))$;
\item The user can specify the output spectral coordinate type (wavelength, freq., energy, or logarithmic) and
a linear spectral coordinate grid (start coordinate of center of first bin and last bin, bin size) ({\bf{SED.an.3.1}});
\item The user can specify the output y-axis type $(F_{\nu}, F_{\lambda}, \log \nu F_{\nu}, \text{etc.}) ({\bf{SED.an.3.2}})$;
\item The tool shall be required to be able to estimate the monochromatic flux at each bin center ({\bf{SED.an.3.3}}). 
For the $k$-th bin, $F(k)$ potentially depends on all the input flux and spectral coordinate values. The user can select 
one of several interpolation algorithms:
	\begin{itemize}
	\item Simple linear interpolation with median: find the input bin coordinates with $x_{ij}$ values $(x_{-}, x_{+})$ closest to and on either side of 
	the target bin $x_k$. If more than one input bin has that value of $x_{ij}$, median those fluxes (for example, multiple different measurements of 
	the magnitude in the V filter done on different dates may be available), leading to $F(x_{-}), F(x_{+})$; then a linear interpolation can be used  
	to find $F(x_k)$. The Kaplan-Meier median is also useful to take upper limits into account and to evaluate the confidence interval on $F(x_k)$ at 
	a given level of confidence from those on the input points taking the interpolation weighting into account; 
	\item Linear interpolation with average: average instead of median in method above;
	\item Smoothing algorithm: specify a smoothing window of spectral coordinate width dx, and a smoothing
	function (support at least boxcar, triangle, gaussian, n-point moving average) with user-supplied parameters. 
	The smoothing is truncated outside the window. Note that an aggregate SED may have spectral coordinate
	regions with very sparse data (e.g. a few bands in the radio and far infrared) where smoothing multiple points 
	together is not desirable because they are far apart, and regions with very dense data (thousands of points
	in the optical with high resolution spectra) where a heavy smoothing might be useful. This is the motivation
	for having a truncation window as well as a smoothing function.
	\end{itemize}
\end{itemize}

\subsection{Fitting of spectral models to a SED and estimation of integrated quantities}
\label{fitspectralmodels}

Another basic requirement of the SED tool shall be the ability to fit the SEDs  ({\bf{SED.an.4}}) with analytical and tabular functions as source models.
The SED analysis tool is required to be able to fit an aggregate SED ({\bf{SED.an.4.1}}) or rebinned SED ({\bf{SED.an.4.2}}) with different spectral models, either already implemented in a 
library of commonly used models or defined by the user either in an analytical way (through a function associated to the analytical expression of 
the mathematical function and written in a chosen programming language) or empirical way (defined by the data contained in a table which can 
be uploaded to the tool). A strong candidate which can provide the core capabilities of the SED fitting and modeling tool is the Sherpa\footnote{Official webpage at the URL http://cxc.harvard.edu/sherpa/}
package. Sherpa, initially developed for the analysis of X-ray data taken by Chandra, today is a mature and general tool designed that can be imported as a module for the Python 
scripting language and is available as a C/C++ library for software development. Although Sherpa seems fit to be employed for the general fitting options described in this section, 
the SED tool shall be required to support other specialized fitting packages providing optimized algorithm for the limited spectral ranges (for example, the SMART and 
PAHFIT\footnote{More details at the URL http://tir.astro.utoledo.edu/jdsmith/research/pahfit.php} which are widely used for fitting and modeling of the mid-infrared and 
far-infrared regions of the SED). Below is a description of the specific requirements for this section of this document.
\begin{itemize}
\item The SED will be approximated by a realization of a model (analytical or algorithmically defined)
\[(s_i, f(s_i)) \rightarrow (M(s), R(s), \text{method}, \vec{p},...)\] where M is the model, R is the range (and grid) 
of values over which the model is compared, `method' is the fitting method (including selection of an optimization 
method and a fit statistic) and $\vec{p}$ is the vector of parameter values. The tool shall identify the parameter values 
which give the optimum value of the fit statistic;
\item The analytical or empirical model M can be selected from a large library of spectral shapes, listed for distinct 
classes of astronomical sources and physical emission mechanisms ({\bf{SED.an.4.3}});
\item M can be specified as an algebraic composition of multiple individual models from the library: 
$(s_i, f(s_i)) \rightarrow (M^{(1)}(s)*M^{(2)}(s), R(s), \text{method}, \vec{p},...)$ where the $*$ operator 
indicates a generic arithmetic combination of two different spectral models, or, more generally, the result 
of the application of any function $F$ on a given model: $(s_i, f(s_i)) \rightarrow F(M(s)), R(s), \text{method}, \vec{p},...)$ ({\bf{SED.an.4.4}});
\item The user can specify the range of spectral coordinates to be considered ({\bf{SED.an.4.5}}), which can be either the whole range, or multiple 
disjoint intervals of spectral coordinates selected by the user: \[(s_i, f(s_i)) \rightarrow (M(s), R^{(1)}(s), R^{(2)}(s), R^{(3)}(s),..., \text{method}, \vec{p},...)\]
These intervals can be provided either by uploading a table containing the extremes of the bins or interactively by the user;
\item The tool shall be able to estimate the goodness-of-fit of the model ({\bf{SED.an.4.6}}) employed letting the user choose among a small number of reference 
fit statistics (for example, a $\chi^{2}$ statistics and some of the specific statistics derived by the maximum likelihood principle;
\item The tool shall be able to evaluate the confidence levels for the parameters ({\bf{SED.an.4.7}});
\item The tool shall support user-defined models ({\bf{SED.an.4.8}}). As already stated above, a fundamental requirement of this
tool shall be to let the user define his own spectral model for the fit analytically. The user could provide a model definition
using a given syntax, i.e. by programmatically defining a function representing the model to be uploaded to the SED analysis tool. A possible 
candidate for the language used for such mechanism of model-definition could be the Python syntax already used in Sherpa; 
\item The tool shall support user-supplied tabular function models ({\bf{SED.an.4.9}}), by uploading a table containing a rebinned SED to be interpolated, in order
to use the interpolated function as spectral model for the fit;
\item This tool shall also support user-defined statistics ({\bf{SED.an.4.10}}). The mechanism for the upload to the tool and the 
required syntax for the user-provided statistics could be the same of the user-provided source models;  
\item Another important requirement of the tool is the ability to calculate integrated fluxes ({\bf{SED.an.4.11}}) in intervals of spectral coordinates
$(R^{(1)}(s), R^{(2)}(s), R^{(3)}(s))$ defined by the user (exploiting  the same mechanism described above for the fit of the 
model in different disjunct intervals of spectral coordinates) from the functional curve generated by the best-fit model of the SED:  
\begin{multline}
(M(s), s, \text{method}, \vec{p},...) \rightarrow \\
\{(R^{(1)}(s), \int M(s)ds, \text{method}, \vec{p},...), R^{(1)}(s))), \\
(R^{(2)}(s), \int M(s)ds, \text{method}, \vec{p},...), R^{(2)}(s))),...\}
\end{multline} 
\item In order to allow easy comparison of the results obtained with literature or published data, the tool shall be required to provide 
a mechanism to accept the spectral intervals defined by the user ({\bf{SED.an.4.12}}), a library of commonly used literature bands, defined 
with different units of measures (for example, the hard, medium, soft, ultrasoft and broad bands available in the Chandra Source Catalog (CSC), or 
the UBVRI bands of Johnson photometry):
\begin{multline}
(M(s), s, \text{method}, \vec{p},...) \rightarrow \\
\{(\text{band}^{(1)}(s), \int(M(s)ds, \text{method}, \vec{p},...), \text{band}^{(1)}(s))),\\
(\text{band}^{(2)}(s), \int(M(s)ds, \text{method}, \vec{p},...), \text{band}^{(2)}(s))),...\}.
\end{multline}
\end{itemize}
%
%
%
%

\subsection{Template fitting}
\label{templatefitting}

The SED analysis tool is required to be able to perform template fitting ({\bf{SED.an.5}}) of an observed (binned or aggregate/interpolated) SED $(s, f(s))$ using
a single template spectrum or a set of synthetic/observed template spectra $(s, t_1(s), t_2(s), t_3(s),...)$ either provided by the user as a set of local files in 
VO-compliant formats that can be ingested by the tool, or retrieved from an online resource. Moreover, this tool shall support some most used local and/or 
online template spectral libraries even if they are not exposed with VO-compliant protocols. In more detail: 
\begin{itemize}
\item The SED template fitting tool shall support simple routine interfaces for some of the most used template model libraries ({\bf{SED.an.5.1}}) (for example, 
the GALAXEV library of evolutionary stellar population synthesis models (Bruzual $\&$ Charlot 2003) for the spectral evolution of stellar 
population);
\item The SED template fitting tool shall also offer user-interfaces for external tools for the simulation of synthetic template
libraries ({\bf{SED.an.5.2}}) as well, providing a mechanism for the seamless exchange of model parameters (from the SED tool to external programs - either installed
on the local disk or accessible through web applications) models and the generated template libraries (from the external programs back to the SED tool). 
For example, the tool should provide a standardized access to the Starlight spectral synthesis code\footnote{Official webpage at the URL http://www.starlight.ufsc.br} and 
the Cloudy code\footnote{Official webpage at the URL http://www.nublado.org/}, allowing the user to interactively choose, without leaving the SED template fitting tool,  
the configuration parameters and to launch the simulations. The single template spectrum generated or the library of synthetic spectra will then
be used as templates for the template fitting;
\item The SED template fitting tool shall also allow the user to modify a library of templates by adding a spectral component ({\bf{SED.an.5.3}}) (for example, a gaussian profile
or a power law) to all templates by specifying the spectral interval of interest and the parameters of the component;
\end{itemize} 
\noindent This tool is also required to handle spatial and spectral distribution of a source to either perform composite spectral fit ({\bf{SED.an.5.4}}) taking into 
account the spatial contribution to the aperture of each point of the SED, or a subtraction of different
components taking into account the aperture corrections (for example, when subtracting to the observed SED multiple components template
spectra associated to either different physical emission mechanisms or type of sources or spatial models). A very basic capability of this tool
shall be the ability to perform  the statistical comparison of one observed SED with a chosen template spectrum and assess reliably it they
are different realizations of the same underlying spectral distribution (for example, through a Kolmogorov-Smirnov test). In more details, the 
requirements of this tool are  described in the following points:
\begin{itemize}
\item The basic functionality of this tool shall be to provide the user with the ability of reading in a library of template spectral models,
and performing template fit on a rebinned or aggregate SED:  $(s, t_1(s), t_2(s), t_3(s),...) \rightarrow (s_i, f(s_i)) \rightarrow (a_1,
a_2, a_3,...)$. The result of such operation is an optimal fitted SED expressed as a combination of the template spectra according  to the
weights $(a_1, a_2, a_3,...)$, the residuals (by subtraction of the optimal fitted SED from the observed SED) and the relative and
absolute composition of the optimal fitted SED in terms of the members of the library of template spectral models used to perform the
fit. The tool is also required to be able to perform the template fitting only on a region of spectral coordinates (specified with the 
mechanism already described above).
\item The tool shall be required to perform composite spectral fit ({\bf{SED.an.5.5}}) in a defined spectral interval, where spatial models
$\text{sp}(x, y)$ for each spectral component (i.e. each template spectrum used)  and for the observed SED are provided in order to
evaluate the spatial contribution to the flux measured in each point. Given the image of the source, the observed SED with spatial information
$(s_i, f(s_i), (x, y))$  can be fitted using as model obtained by combining multiple single models associated to components with both a
spectral and spatial distribution: 
\begin{multline}
(s, t_1(s), \text{sp}^{(1)}(x, y))*(s, t_2(s), \text{sp}^{(2)}(x, y))*(s, t_3(s), \text{sp}^{(3)}(x, y))*...) \\
\leftrightarrow (s_i, f(s_i), (x, y))
\end{multline}
\item The tool shall also be required to use the spatial information about the apertures available for the observed SED and
templates spectra to correct the observed flux for aperture effect ({\bf{SED.an.5.6}}): $(s_i, f(s_i)) \rightarrow (s_i, f_c(s_i))$, and perform the subtraction
from the observed SED of the template SEDs from the observed SED of template reference spectra associated to different components.
\end{itemize}

\subsection{Population analysis tool}
\label{populationanalysis}

The SED analysis tool is also required to be able to generate {\bf statistical SEDs} from a sample of M rebinned or aggregate SEDs from
different sources ({\bf{SED.an.6}}): 

\begin{equation}
\{(s, i_1(s)), (s, i_2(s)), (s, i_3(s)),...\} \rightarrow (s, s_t(s)) \text{where } i = \{1,..., \text{M}\}
\end{equation}

\noindent where $s_t(s)$ is a SED obtained as, for example, the mean, median or percentile of the parent sample of SEDs. These SEDs may be normalized to
the flux of a given spectral point $f(\tilde{s})$ or to a bolometric or integrated flux $f(A) = \int_{A}f(s)ds$ obtained using a reference or user-supplied interval of spectral
coordinates, or band: $(s, i(s)) \rightarrow (s, i_n(s)) \text{where } i_n(s) = i(s)/f_{A}$. This tool is also required to be capable of evaluating statistical quantities ({\bf{SED.an.6.1}}) for each 
SED of a given sample, in the whole spectral range covered by the SEDs or inside a given spectral interval. For example, the tools shall be able to evaluate the coordinates along 
the spectral and flux axes of the minima, maxima and inflection points in every single SED of the sample, and save such data in the VOtable format:
\begin{multline}
\{(s, i_1(s)), (s, i_2(s)), (s, i_3(s)),...\} \rightarrow \\
\{(s_1^{(min_1)}, f_1^{(min_1)}(s)), (s_1^{(min_2)}, f_1^{(min_2)}(s)),..., (s_1^{(max_1)}, f_2^{(max_1)}(s)), (s_1^{(max_2)}, f_1^{(max_2)}(s)),..., \\
(s_1^{(flex_1)}, f_1^{(flex_1)}(s)), (s_1^{(flex_2)}, f_2^{(flex_2)}(s)),..\}
\end{multline}
The SED tool shall also be able to allow the user to derive few simple descriptive statistical quantities for selected
spectral regions for each SED belonging to the sample ({\bf{SED.an.6.2}}), as the FWHM of a given {\it bump}, the 
linear regression parameters of a given segment of the interpolated SED inside the spectral interval
$R(s)$ and/or for flux values inside an interval $[f_{-}, f_{+}]$, etc.). An additional requirement of this tool 
is the ability to evaluate integrated quantities ({\bf{SED.an.6.3}}) (integrated flux, flux ration, i.e. colors, etc.) for all SEDs of 
the sample for a set of spectral intervals ($R_1(s)$, $R_2(s)$,...) or bands ($A$, $B$,...) specified with the usual mechanism. 
In general:
\begin{multline}
\{(s, i_1(s)), (s, i_2(s)), (s, i_3(s)),...\} \rightarrow \{\text{FWHM}_1(\text{bump}), \text{FWHM}_2(\text{bump}),...\}\\
\{(s, i_1(s)), (s, i_2(s)), (s, i_3(s)),...\} \rightarrow \{(\alpha, \beta)_1(R(s)), \alpha, \beta)_1([f_{-}, f_{+}]), \\
(\alpha, \beta)_2(R(s)), \alpha, \beta)_2([f_{-}, f_{+}]),...\}\\
\{(s, i_1(s)), (s, i_2(s)), (s, i_3(s)),...\} \rightarrow \{(f_1(S_1), f_1(S_2), f_1(S_1)/f_1(S_2)), \\
(f_2(S_1), f_2(S_2), f_2(S_1)/f_2(S_2)),... \}
\end{multline}

\subsubsection{Classification of SEDs}
\label{classificationofseds}

From a general statistical standpoint, classification can be stated as follows:  a given training dataset $\{(x_1, y_1), (x_2, y_2),..., (x_M, y_M)\}$ produce a
classifier $h\text{: X}\rightarrow\text{Y}$ that maps any object $x \in\text{X}$ to its classification label $y \in\text{Y}$ defined by some unknown mapping function 
$g\text{: X}\rightarrow\Omega$. The SED tool shall be required to provide classification capabilities for a sample of SEDs ({\bf{SED.an.7}}). In more details,   
the analysis tool shall provide two different mechanisms to recognize similarities among distinct members of a given SEDs population ({\bf{SED.an.7.1}}) (clustering) and, given different 
classes of SEDs, assign new SEDs to one of such classes on the basis of their shape ({\bf{SED.an.7.2}}). With a sample of $M$ SEDs $\{(s, i_1(s)), (s, i_2(s)), (s, i_3(s)),...\}$ 
defined in the same spectral range, the SED population analysis tool is required to be able to:
\begin{enumerate}
\item group SEDs into different classes $C_i$ according to their shapes over the whole spectral domain or in a given interval $[f_{-}, f_{+}]$ (defined by the user with the same mechanism
described in the previous sections):
	\begin{equation}
	\{(s, i_1(s)), (s, i_2(s)), (s, i_3(s)),...\} \rightarrow \{C_1, C_2,...\}
	\end{equation}
The tool shall also perform the grouping on the basis of a set of parameters associated to models of part or the entire SEDs obtained through 
any of the analysis steps available in the SED analysis tool. For example, the grouping could be performed on the basis of a set of shape parameters for a
given spectral interval: 
	\begin{equation} 
	\{(\text{FWHM}_1(\text{bump}), \alpha_1, \beta_1,...), (\text{FWHM}_2(\text{bump}), \alpha_2, \beta_2,...),..')\} \rightarrow \{C_1, C_2,...\}
	\end{equation}
Several distinct statistical methods shall be available to extract a set of classes from a population of SEDs (Fisher's matrix correlations, )	
\item classify a new SED $(s, \hat{i}(s))$ by associating it to one of the classes $\{C_1, C_2,...\}$. Such classes can be defined either on the basis of a set 
of 'prototypical' SED models (assigned with the same mechanism used for the retrieval and definition of the templates SEDs discussed for the SED template 
fitting tool in section \ref{templatefitting}):
	\begin{equation}
	\{C_1, C_2,...\} \leftrightarrow (s, \hat{i}(s)) \rightarrow \hat{C}
	\end{equation}
or by constraining the values of a set of parameters $\{\text{FWHM}, \alpha, \beta,...\}$ associated to a possible parametrization of a distinct spectral segment or of 
the whole SED:
	\begin{equation}
	\{[\text{FWHM}_{-}^{(i)}, \text{FWHM}_{+}^{(i)}],  [\alpha_{-}^{(i)}, \alpha_{+}^{(i)}], [\beta_{-}^{(i)}, \beta_{+}^{(i)}],...\} \rightarrow C_i 
	\end{equation}
so that it is possible to use the values of the same parameters for the new SED: 
	\begin{equation}
	(s, \hat{i}(s)) \rightarrow (\hat{\text{FWHM}}, \hat{\alpha}, \hat{\beta},...) 
	\end{equation}
	\begin{equation}
	\{(\text{FWHM}_1(\text{bump}), \alpha_1, \beta_1,...), (\text{FWHM}_2(\text{bump}), \alpha_2, \beta_2,...),..')\} \leftrightarrow (\hat{\text{FWHM}}, \hat{\alpha}, \hat{\beta},...) \rightarrow \hat{C}
	\end{equation}
\end{enumerate}
These two different tasks represent two different statistical paradigms: the determination of classes of similar objects with no {\it a priori} assumption is an example
of an unsupervised operation, while classification is a more canonical supervised operation. The SED population analysis tool shall provide multiple statistical methods
to perform both tasks. An example of a simple statistical technique that can be used to obtain an unsupervised classification through the clustering of a population is the 
'k-means' method, which can be applied either on the whole SEDs (represented as $2xN$ dimensional vectors, where N is the total number of photometric points and/or 
segment) or on a set of parameters associated to the description of whole or part of the SEDs. On the other side, classification can be achieved using several methods, for 
example the Fisher's linear discriminant analysis, Support Vector Machine, neural networks, the already discussed Kolmogorov-Smirnov test (\ref{templatefitting}), etc.

\subsection{Astrophysics and Cosmology tools}
\label{astrophysicsand}

In addition to the analysis capabilities described so far, the SED analysis tool shall be required to support  two different types of physical analysis ({\bf{SED.an.8}}), namely the 
determination of the extinction based on extinction laws ({\bf{SED.an.8.1}}) in multiple spectral regions, the evaluation of the effect on the observed SED of the redshift $z$ and the 
underlying cosmological model ({\bf{SED.an.8.2}}), which is considered by default as the WMAP-5 cosmological model, the evaluation of the bolometric luminosity for the 
source ({\bf{SED.an.8.3}}) from the measured flux and an estimate of the distance provided by the user, and the ability for the user to change the default values of the cosmological
parameters ({\bf{SED.an.8.4}}). More details about these capabilities are given in the following points:
\begin{itemize}
\item A fundamental requirement of the SED tool is the ability to calculate the extinction and perform the ''de-reddening'' of the SED by applying standard 
or user-provided extinction laws  $\{(s_A, e(s_A)), (s_B, e(s_B)), (s_C, e(s_C)),...\}$ in different bands (for example, gas column density for X-ray wavelengths 
and ultraviolet/optical/infrared dust column density) ({\bf{SED.an.8.1.1}}). The tool shall be required to let the user choose from a collection of standard extinction 
laws already available as different libraries offered by the same tool ({\bf{SED.an.8.1.2}}), or to define his own extinction law by uploading a file containing a tabular 
description of the extinction curve in a defined spectral region ({\bf{SED.an.8.1.3}}). Moreover, the tool shall be required to support the evaluation of the extinction 
using composite models with redshift dependence for each component of the model: $(s_A, e(s_A, z))*(s_B, e(s_B, z))$ ({\bf{SED.an.8.1.4}});
\item This tool shall be required to be able to perform conversion of the SED between different reference frames of the observed SED ({\bf{SED.an.8.2}}), for a given 
set (defined by the user) of parameters of  the standard cosmological model: 
$(s_i, f(s_i), z_1) \rightarrow (s_i(z_2), f(s_i(z_2)), z_2)$;
\item The SED tool shall also be required to perform conversion from monochromatic flux or defined for a given interval of spectral coordinates to luminosity 
({\bf{SED.an.8.3}}) using a value of the distance of the source $d$ provided by the user:
\begin{equation}
(s, f(s), d) \rightarrow \text{L}(d)
\end{equation}
\noindent The tool will also be able to calculate the bolometric luminosity of the source associated to a given rebinned SED or model for the observed SED.
\item The user may set the standard cosmological parameters $H_{0}, \Omega_0, \Omega_m, \Lambda$ ({\bf{SED.an.8.4}})
The WMAP 5-year values shall be used as the default.
\end{itemize}

\subsection{Convolution tool}
\label{convolutiontool}

The SED analysis tool shall be required to support the convolution of a given SED with a generic function of the spectral coordinates ({\bf{SED.an.9}}). 
In general, the convolution of a given SED $(s, f(s))$ with a generic function $g(s)$ is defined as:

\begin{equation}
(f\star g)(s) = \int_{-\infty}^{\infty}f(\tau)g(s-\tau)d\tau
\end{equation}

\noindent The function $g(s)$ can be either analytically defined (boxcar function, gaussian, triangular, etc.), or defined by its tabular representation (for example, 
an instrumental resolution profile, which is a simple special case of a SED). The tool shall provide the user with a set of common analytical functions and common 
instrumental profiles. A further requirement of this tool will be the capability of supporting also analytical or tabular user-defined convolving functions (with a mechanism 
similar to what described for user-supplied model sources, methods, extinction laws,... in the previous paragraphs). It is also worth stressing that, even if some of the 
smoothing capabilities described in paragraph \ref{convertanaggregatesed} of this document can be regarded as particular cases of convolution, the two functionalities 
are kept distinct. 

\section{SED visualization and editing}
\label{sedvisualizationandediting}

\subsection{SED visualization}
\label{sedvisualization}

The SED display tool shall be required to be able to display one or more aggregate or rebinned SEDs ({\bf{SED.vis.1}}), SED derived from spectral model, template SED letting the user decide 
a number of visualization options:
\begin{itemize}
\item The tool should allow the user to plot the SED as 'generic flux' versus spectral coordinates and interactively convert between multiple representations of the SED ({\bf{SED.vis.1.1}})
(at least $f_{\lambda}$, $f_{\nu}$, $\nu f_{\nu}$, photon flux for the y-axis; wavelength, energy, frequency for the x-axis) with a variety of units (SI, cgs and $L_{\odot}$, $M_{\odot}/\text{yr}$, 
MJy$/$yr; \AA, $\mu$m, nm, cm, erg, keV, Hz, GHz,...);
\item The tool shall support the interactive selection of a spectral region and the re-plotting of the SED in this region ({\bf{SED.vis.1.2}}), as well as zooming in and out of the SED ({\bf{SED.vis.1.3}}). 
This shall be possible using either the pointing device for a 'quick and dirty' selection of a given spectral region, or assigning numeric values to the extremes of the spectral region
in a graphical window for a more accurate selection. 
\end{itemize} 

\subsection{Interaction with the SED}
\label{interactionwith}

The SED display tool shall be required to support the interaction of the user with the SED plot in multiple ways, for both single photometric points and spectral segments.
The range and scope of the interactivity offered by the SED visualization tool shall encompass multiple aspects of the analysis and production of the displayed 
SED. In general, this tool shall allow the user to interactively perform all the possible analysis steps of the SED during a work session, display the result of each operation and 
revert the SED and all derivate data to any previous state during the session. The process of building a SED often requires the inspection of the data by the user; for this reason, 
the SED visualization tool shall also support a mechanism to interact with every single data element constituting the SED (photometric points and spectral segments), or 
to interactively define subsets of data elements, consisting of different points or entire spectral regions, flag them for deletion or interactively modify their positions in the 
spectral plot. The tool shall support direct operations on a selected subset of data elements 
(i.e. averaging multiple fluxes or spectral segments, simple mathematics - sum, subtraction, product and division - on the members of the subset, aperture correction with a 
given spatial model of the source, etc.) and display as new data elements the resulting points or spectral segments, allowing the user to retain them and saving a new version 
of the SED or to discard them. In a generic interactive session, the user shall be able to:
\begin{itemize}
\item modify the SED structure (temporarily or permanently) by flagging single components of the SED ({\bf{SED.vis.1.4}}), decide to save the modified SED to a different VO-compliant file;
\item shift points or set of points or entire spectral segments along both the x and y axes by ``clicking and dragging" ({\bf{SED.vis.1.5}});
\item adjust the data (for example, changing the curvature of a spectral segments, modifying the error bars, changing the upper limits) by ``clicking and dragging" ({\bf{SED.vis.1.6}});
\item interactively inspect the metadata ({\bf{SED.vis.1.7}}) (time of the observation, aperture, photometric system, reference in the literature, name of the PI, parameters of the reduction of the data,...) 
associated to each single photometric point or spectral segment composing the SED by hovering the pointer over any data element  ({\bf{SED.vis.1.7.1}}) or set of data elements with 
common origin, be able to save such metadata to a file ({\bf{SED.vis.1.7.2}}) and/or display any online available information related to the data in an external web browser ({\bf{SED.vis.1.7.3}});
\item export the modified version of the SED or any section defined in spectral coordinates of it to a VO-compliant file ({\bf{SED.vis.1.8}}), or plot it as a new SED displayed in a new window ({\bf{SED.vis.1.9}});
\item perform simple mathematical operations on the SED or an interactively selected region of the SED ({\bf{SED.vis.1.10}}) (for example, adding, subtracting, multiplying or dividing x or y values by a constant);
\item save the plot in different graphical formats  ({\bf{SED.vis.1.11}});
\end{itemize}

\subsection{Session report and CLI scripting}
\label{sessionreport}

The SED tool, alongside with the interactive graphical interface, shall support a mechanism that can record all the operations ({\bf{SED.vis.2}}) performed during a work session by the user
and let the user repeat such analysis and reductions steps from in a not interactive way. For this reasons, the SED analysis shall provide two different types of files produced for any interactive 
session: a session report ({\bf{SED.vis.2.1}}) and a workflow script ({\bf{SED.vis.2.3}}), the former being a description as accurate as possible of the workflow operations, and the latter a script 
containing a collection of CLI commands associated to each operation of the workflow. The session report should contain a complete list of the operations (and their results) performed by the 
user during the work session and, in particular, should keep record of: 
\begin{enumerate}
\item all changes to the visualization options;
\item all interactive changes to the values of single photometric points and spectral segments or user-defined spectral intervals of the SED; 
\item the corresponding changes in the SED structure (in particular, when the modified SED has not been saved to a different file);
\item the results (for example, best fit parameters or integrated fluxes for a set of filters, coefficient of the template fitting, etc.) of the interactive 
analysis steps (aperture correction, fitting, interpolation, arithmetic operations, population related statistics, template fitting, smoothing, 
convolution, etc.) carried out during the session; 
\item the configuration parameters and metadata of each analysis step performed by the user;
\end{enumerate}
A stripped-down version of the session report described above ({\bf{SED.vis.2.2}}), containing only the description of the analysis steps performed
during an interactive work session by the user and the references to saved files containing the SED data should supported by the SED tool as an 
early requirements. This simplified session report would not mention any interactive change in the visualization of the plots. The SED tool shall be 
required to provide a mechanism for producing a workflow script ({\bf{SED.vis.2.3}}) in text format which can be understood (and modified) by the 
user and re-used, in batch mode, to replicate not interactively the analysis workflow from the CLI.

\subsection{Tabular data visualization}
\label{tabulardata}

The SED display tool shall be required to provide simple visualization capabilities for the different quantities evaluated during the SED analysis ({\bf{SED.vis.3}}) 
(integrated fluxes, luminosities, magnitudes, colors, hardness ratios, spectral indexes, extinctions, redshifts, etc.). It shall be able to produce simple 2-d scatter plots ({\bf{SED.vis.3.1}})
and histograms ({\bf{SED.vis.3.2}}), with the interactive choice of the parameters and coordinate ranges (in other words, a stripped-down version of the TOPCAT 2-d scatter plot
and histogram capabilities). This tool shall also be required to produce simple graphical representations of the spatial information (spatial models or apertures) associated to a given
aggregate SED, if available in the metadata of the file ({\bf{SED.vis.3.3}}).

\section{Other SED tools}
\label{othersedtools}

A comparison table for some of the tools that are available today to build SEDs from archival data and perform analysis is shown below. The tools compared are the following: 
VO SED Analyzer\footnote{Official webpage at the URL http://www.laeff.inta.es/svo/theory/vosa/}, VOSED\footnote{Official webpage at the URL http://sdc.laeff.inta.es/vosed}, 
VOSpec\footnote{Web application can be found at the URL http://esavo.esa.int/vospec/ and the user manual at the URL http://esavo.esac.esa.int/VOSpecManual/}, 
SPLAT\footnote{User manual at the webpage http://star-www.dur.ac.uk/~pdraper/splat/sun243.htx/sun243.html} and ASDC SED Builder\footnote{Web application can be 
found at the URL http://tools.asdc.asi.it/SED/}. The capabilities of each of these tools have been evaluated according to the break-up of the requirements used in this document. 
So, every single tool will be ranked against the requirements described in this document using one of the following symbols:
\begin{itemize}
\item $< $: the tool has capabilities less extended than those of the requirement SED.xxx.xx described in this document; 
\item $> $: the tool has capabilities more extended than those of the requirement SED.xxx.xx described in this document; 
\item $= $: the tool has the same capabilities of the requirement SED.xxx.xx described in this document;
\item $\neq $: the tool has different capabilities relative to the requirement SED.xxx.xx described in this document;
\item $-$: the tool has no capabilities for the requirement SED.xxx.xx described in this document;
\item ?: it was not possible to establish if the tool has the capabilities for the requirements SED.xxx.xx described in this document;
\end{itemize}
\noindent These symbols can be combined (for example, $<<$ means ``capabilities much more limited that those described in this document", or
$<\neq$ means ``capabilities less extended than those described in this document and partially different").

\begin{sidewaystable}
\caption{Comparison table of available tools for SED creation, analysis and visualization.}
\centering
{\footnotesize
\begin{tabular}{l c c c c c c c c c c c c c c c c c c}
\hline\hline
	& \multicolumn{6}{|c|}{SED Builder} & \multicolumn{9}{|c|}{SED Analysis} & \multicolumn{3}{|c|}{SED Visualization} \\  
Tool & Bui.1 & Bui.2 & Bui.3 & Bui.4 & Bui.5 & Bui.6 & An.1 & An.2 & An.3 & An.4 & An.5 & An.6 & An.7 & An.8 & An.9 & Vis.1 & Vis.2 & Vis.3\\ 
\hline
	&       &       &       &        &       &       &       &       &       &       &       &       &       &       &       &       &       &      \\ 
VOSA &  {\Large $<$}     &  {\Large $=$}     &  {\Large $<$}     &  {\Large  -}     &   {\Large ?}    &   {\Large $<$}    &  {\Large $<<$}   & {\Large  -}  &   {\Large ?}    &  {\Large  $<<$}    &  {\Large $=$}     &  {\Large  -}    &   {\Large  -}   &   {\Large  -}   &   {\Large -}    &  {\Large -}  &   {\Large $<<$}    &  {\Large -}    \\ 
	&       &       &       &        &       &       &       &       &       &       &       &       &       &       &       &       &       &      \\ 
VOSED &  {\Large $=$}     &  {\Large $=$}     &  {\Large  -}    &  {\Large  -}     &    {\Large ?}  &  {\Large $=$}   &  \multicolumn{9}{|c|}{VOSpec}   &    \multicolumn{3}{|c|}{VOSpec}      \\
	&       &       &       &        &       &       &       &       &       &       &       &       &       &       &       &       &       &      \\ 
VOSpec &     \multicolumn{2}{|c|}{VOSED}  & {\Large -}  &    {\Large $=$}     &  {\Large ?}   &  {\Large $<$}   &  {\Large $=$}  &  {\Large $<$}  &  {\Large -}  &  {\Large $<$} &   {\Large $=$}    &   {\Large $<$}    &  {\Large -}  &   {\Large $<<$}    &  {\Large $<$} &  {\Large -}  &  {\Large -}  &  {\Large $=$}      \\ 
	&       &       &       &        &       &       &       &       &       &       &       &       &       &       &       &       &       &      \\ 
SPLAT &   {\Large $<=$}    &   {\Large $<=$}     &  {\Large -} & {\Large $=$} &  {\Large ?} &   {\Large $<<$} & {\Large $=$} & {\Large ?} &  {\Large -} &  {\Large $<<$} & {\Large $<\neq$} &  {\Large -} & {\Large -} &  {\Large $<$} &  {\Large -} &   {\Large $<\neq$}    &  {\Large  -}    &  {\Large  -}   \\
	&       &       &       &        &       &       &       &       &       &       &       &       &       &       &       &       &       &      \\ 
ASDC SED Builder &   {\Large $<<$}    &   {\Large $<<$}  &  {\Large -}  & {\Large -} & {\Large ?} & {\Large $<$} & {\Large $=$} & {\Large ?} & {\Large  -} &  {\Large $<$} &    {\Large $<\neq$}  & {\Large  -} & {\Large  -}    &   {\Large $<$}    &  {\Large -}  &  {\Large $<<$}  &  {\Large -}  & {\Large -}    \\
	&       &       &       &        &       &       &       &       &       &       &       &       &       &       &       &       &       &      \\ 
\end{tabular}
}
\label{table:comparison}
\end{sidewaystable}

\noindent In general, each tool addresses only partially the requirements described in this document, and none of them seems to clearly distinguish
between the two basically different types of SED (aggregate and rebinned) indicated here. Synthetic comments for each of the five tools 
compared in the table \ref{table:comparison} can be found below:

\begin{itemize}
\item VOSA: in general, using this tool can be confusing and impractical for a number of reasons (the authors claim that it is still 
in development though). While fairly satisfactory for the construction of the SED from data available through VO-compliant services (it is the only 
tool to support the access to precomputed stellar SEDs), it falls short of providing almost all the other requirements for the analysis and 
visualization of the SED (except for the template fitting capability, with few libraries that can be accessed remotely); 
\item VOSED: the capabilities of this tool concerning the construction of SED from distinct data elements are good for the data resources exposed
through VO protocols and the major archival data provider  (though SAMP integration and the ability to support precomputed SEDs are entirely missing). 
The analysis and visualization of the SEDs are referred to the companion interactive client, VOSpec; 
\item VOSpec: the construction capabilities are almost completely missing, since VOSpec relies on VOSED for construction of the SED from data elements
accessible through VO protocols, except for the loading of local data. The analysis capabilities of this tool are good for what fitting and modeling are 
concerned (even if the submission of user-defined models and statistics is not allowed), while template fitting is possible only for very simple spectral 
templates, and the whole population libraries are not allowed. No population analysis, classification capabilities and convolution tool are provided. 
The visualization of the SED is clear and many options are offered to customize the appearances of the plot, together with the ability to perform simple 
arithmetic and geometrical operations on the displayed SED; on the other hand, the metadata associated to the different data elements cannot be accessed 
interactively and no session report capability is supported. An interface for the SAMP protocol easily allows the exchange of SED and generic data with other tools; 
\item SPLAT: in general, SPLAT offers a functional interface for the retrieval of spectral segments exposed through VO protocols and is able to inspect the
metadata associated to each data element from the same window used to load the data. The analysis of the spectral data is somewhat limited; for example, 
support for only polynomial fitting and no template fitting capability at all are offered. Spectral line profile fitting capability is available, with1 the most common functions used as
profiles (it should be noticed that line profile fitting is not a specific requirement of this document for the SED tool even if it can be performed as a specific
case of the more general fitting requirements). No classification capability is available, while the it is possible to modify a spectral segment by changing 
its redshift. On the other hand, the tool offers fairly sophisticated options for tweaking the visualization of the spectra, and supports the SAMP protocol for
application interoperability. In any case, it should be stressed that SPLAT has been designed to handle spectral data and does not support loading, retrieval, 
analysis and visualization of photometric quantities like fluxes;
\item ASDC SED Builder: a weak point of this tool is that, apparently, it completely lacks any support of the VO protocol for data retrieval and the absence of an 
interface for the SAMP interconnection. On the other side, the interface for the retrieval of both spectroscopic and photometric data from the ASI archive
and other services (like Vizier) is simple and powerful. The analysis capabilities are fairly complete, encompassing almost all the topics touched in this
document except for the explicit requirement for the convolution and the population analysis. This tool does not allow user supplied source models 
and statistics, and the template fitting is limited to few SED template for certain types of sources; at the same time, it supports direct access to several instrumental 
sensitivities profile (this is not a requirement for the SED tool described in this document). Visualization offers few options to customized the plotting 
of the SED but no session report, interactive inspection of metadata and visualization of tabular data are supported.  
\end{itemize}

\section{Breakdown of scientific priorities}

\begin{table}
\caption{Hierarchical break-down of the requirements and proposed prioritization of the deliverables of the ``SED builder" section of the document.}
\centering
\begin{tabular}{r r c c c c c c}
\hline
Task & Sub-task & Sub-sub-task & Description & Year 1 & Year 2 & Later... \\
\hline
SED.bui.1		&				&					&	{\scriptsize Ingest local data}			&	{\large X} 		& 	 			&				\\
$\searrow$	& 	SED.bui.1.1	& 	 				&	{\scriptsize Ingest/convert tables}		&	{\large X} 		& 	 			&				\\
		$|$	& 	SED.bui.1.2	&	 				&	{\scriptsize Ingest/convert spectrum}		& 	{\large X} 		&				&				\\
		$|$	& 	SED.bui.1.3	&	 				& 	{\scriptsize Ingest spectral datacube}	&	 			&	{\large X}		&				\\
		$|$	& 	$\searrow$	&	SED.bui.1.3.1	 	&	{\scriptsize Slit spectroscopy}			&	 			& 	{\large X} 		&				\\
		$|$	& 			$|$	& 	SED.bui.1.3.2	 	&	{\scriptsize Slitless spectroscopy} 		&	 			& 	{\large X} 		&				\\
		$|$	&			$|$	& 	SED.bui.1.3.3	 	&	{\scriptsize Integral field spectroscopy}	&	 			& 	{\large X} 		&				\\
		$|$	&			$|$	&	SED.bui.1.3.4	 	&	{\scriptsize Fiber spectroscopy}			&	 			& 	{\large X} 		&				\\
		$|$	&			$|$	&	SED.bui.1.3.5	 	&	{\scriptsize Photometry-based spec.}	&	 			& 	{\large X} 		&				\\
		$|$	&			$|$	&	SED.bui.1.3.6	 	&	{\scriptsize Time-dependent spec.}		&	 			& 	{\large X} 		&				\\
		$|$	&			$|$	&	SED.bui.1.3.7	 	&	{\scriptsize Convolve with phot. filter}	&	 			& 	{\large X} 		&				\\
		$|$	&	SED.bui.1.4	&	 				&	{\scriptsize Ingest theor. spectrum}		& 	 			&	{\large X}		&				\\
		$|$	& 			$|$	& 	SED.bui.1.4.1	 	&	{\scriptsize Spectrum from model}		&	 			& 	{\large X} 		&				\\
		$|$	&			$|$	& 	SED.bui.1.4.2	 	&	{\scriptsize Grid of spectra from model}	&	 			& 	 			&	{\large X}		\\
		$|$	&	SED.bui.1.5	&	 				&	{\scriptsize Extract phot. from image}		& 	 			&	{\large X}		&				\\
		$|$	&	$\searrow$	&	SED.bui.1.5.1	 	&	{\scriptsize Back. and error images}		&	 			& 	{\large X} 		&				\\
		$|$	& 			$|$	& 	SED.bui.1.5.2	 	&	{\scriptsize User-supplied region}		&	 			& 	{\large X}	 	&				\\
		$|$	&			$|$	& 	SED.bui.1.5.3	 	&	{\scriptsize Regions for image and back.}	&	 			& 			 	&	{\large X}		\\
		$|$	&			$|$	&	SED.bui.1.5.4	 	&	{\scriptsize Measure flux from region}	&	 			& 			 	&	{\large X}		\\
		$|$	&			$|$	&	SED.bui.1.5.5	 	&	{\scriptsize Aperture correction}		&	 			& 			 	&	{\large X}		\\
		$|$	&			$|$	&	SED.bui.1.5.6	 	&	{\scriptsize Spectrum from image}		&	 			& 			 	&	{\large X}		\\
SED.bui.2		&				&					&	{\scriptsize Ingest VO-published data}	&			 	& 	{\large X}	 	&				\\
$\searrow$	& 	SED.bui.2.1	& 	 				&	{\scriptsize Simple Image Access}		&	 			& 	{\large X}		&				\\
		$|$	& 	SED.bui.2.2	&	 				&	{\scriptsize Simple Spectral Access}		& 	 			&	{\large X}		&				\\
		$|$	& 	SED.bui.2.3	&	 				&	{\scriptsize Spectral Lines Access}		& 	 			&	{\large X}		&				\\
		$|$	& 	SED.bui.2.4	&	 				&	{\scriptsize Tabular (data) Access}		& 	 			&				&				\\
SED.bui.3		&				&					&	{\scriptsize Precomputed SEDs}		&	{\large X} 		& 			 	&				\\
$\searrow$	& 	SED.bui.3.1	& 	 				&	{\scriptsize Science collaborations}		&	 			& 	{\large X}		&				\\
		$|$	& 	SED.bui.3.2	&	 				&	{\scriptsize Data archives services}		& 	{\large X} 		&				&				\\
SED.bui.4		&				&					&	{\scriptsize SAMP interface}			&	{\large X}	 	& 			 	&				\\
SED.bui.5		&				&					&	{\scriptsize Usage metrics}			&	{\large X} 		& 			 	&				\\
SED.bui.6		&				&					&	{\scriptsize Assemble aggregate SEDs}	&	{\large X} 		& 			 	&				\\
SED.bui.7		&				&					&	{\scriptsize Aperture corrections}		&	{\large X} 		& 			 	&				\\
$\searrow$	& 	SED.bui.7.1	& 	 				&	{\scriptsize Simple aperture corr.}		&	{\large X} 		& 				&				\\
		$|$	& 	SED.bui.7.2	&	 				&	{\scriptsize Model-based aperture corr.}	& 	 			&				&	{\large X}		\\

\end{tabular}
\label{table:prioritizationSEDbui}
\end{table}

\begin{table}
\caption{Hierarchical break-down of the requirements and proposed prioritization of the deliverables for the ``SED analysis" section of the document.}
\centering
\begin{tabular}{r r c c c c c c}
\hline\hline
Task & Sub-task & Sub-sub-task & Description & Year 1 & Year 2 & Later... \\
\hline
SED.an.1		&				&					&	{\scriptsize Read/convert aggregate SED}&	 {\large X}		& 	 			&				\\
SED.an.2		&				&					&	{\scriptsize Read rebinned SED}		&	 			& 	{\large X} 		&				\\
SED.an.3		&				&					&	{\scriptsize Convert from agg.to rebinn.}	&	 			& 	{\large X} 		&				\\
$\searrow$	& 	SED.an.3.1	& 	 				&	{\scriptsize Choose spectral coord. grid}	&	 			& 	{\large X} 		&				\\
		$|$	& 	SED.an.3.2	&	 				&	{\scriptsize Choose units of measure}	& 			 	&	{\large X}		&				\\
		$|$	& 	SED.an.3.3	&	 				&	{\scriptsize Measure bin fluxes}			& 	 			&	{\large X}		&				\\
SED.an.4		&				&					&	{\scriptsize Fitting of SEDs}			&	 {\large X}		& 			 	&				\\
$\searrow$	& 	SED.an.4.1	& 	 				&	{\scriptsize Fit aggregate SEDs}		&	 {\large X}		& 				&				\\
		$|$	& 	SED.an.4.2	&	 				&	{\scriptsize Fit rebinned SEDs}			& 	 			&	{\large X}		&				\\
		$|$	& 	SED.an.4.3	&	 				&	{\scriptsize Library of template models}	& 	 {\large X}		&				&				\\
		$|$	& 	SED.an.4.4	& 	 				&	{\scriptsize Composed models}			&	 {\large X}		& 			 	&				\\
		$|$	& 	SED.an.4.5	&	 				&	{\scriptsize Specify spectral interval}		& 	 {\large X}		&				&				\\
		$|$	& 	SED.an.4.6	&	 				&	{\scriptsize Estimate goodness-of-fit}		& 	 {\large X}		&				&				\\
		$|$	& 	SED.an.4.7 	&	 				&	{\scriptsize Estimate confidence lev.}	& 	 {\large X}		&				&				\\
		$|$	& 	SED.an.4.8	&	 				&	{\scriptsize User-def. model funct.}		& 	 			&	{\large X}		&				\\
		$|$	& 	SED.an.4.9	&	 				&	{\scriptsize User-def. tabular models}	& 	 			&	{\large X}		&				\\
		$|$	& 	SED.an.4.10	& 	 				&	{\scriptsize User-def. statistics}			&	 			&		 	 	&	{\large X}		\\
		$|$	& 	SED.an.4.11	&	 				&	{\scriptsize Estimate integr. fluxes}		& 	 			&	{\large X}		&				\\
		$|$	& 	SED.an.4.12	&	 				&	{\scriptsize Estimate fluxes in ref. filters}	& 	 			&				&	{\large X}		\\
SED.an.5		&				&					&	{\scriptsize Template fitting}			&	 			& 	{\large X} 		&				\\
$\searrow$	& 	SED.an.5.1	& 	 				&	{\scriptsize Routine interf. for libraries}	&	 			& 	{\large X} 		&				\\
		$|$	& 	SED.an.5.2	&	 				&	{\scriptsize Interf. to synthetic libraries} 	& 	 			&				&	{\large X}		\\
		$|$	& 	SED.an.5.3	&	 				&	{\scriptsize Add comp. to libraries}		& 			 	&				&	{\large X}		\\
		$|$	& 	SED.an.5.4	& 	 				&	{\scriptsize Spatial-spectral fit} 			&	 			& 			 	&	{\large X}		\\
		$|$	& 	SED.an.5.5	&	 				&	{\scriptsize Spatial-spectral fit in interv.}	& 	 			&				&	{\large X}		\\
		$|$	& 	SED.an.5.6	&	 				&	{\scriptsize Evaluate aperture correct.}	& 	 			&				&	{\large X}		\\
SED.an.6		&				&					&	{\scriptsize Population analysis}		&	 			& 	{\large X} 		&				\\
$\searrow$	& 	SED.an.6.1	& 	 				&	{\scriptsize Evaluate pop. statistics}		&	 			& 	{\large X} 		&				\\
		$|$	& 	SED.an.6.2	&	 				&	{\scriptsize Eval. pop. statistics in interv.}	& 	 			&				&	{\large X}		\\
		$|$	& 	SED.an.6.3	&	 				&	{\scriptsize Eval. integrated quantities}	& 	 			&	{\large X}		&				\\
SED.an.7		&				&					&	{\scriptsize SEDs classification}		&			 	& 			 	&	{\large X}		\\
$\searrow$	& 	SED.an.7.1	& 	 				&	{\scriptsize Clustering of SEDs}			&	 			& 			 	&	{\large X}		\\
		$|$	& 	SED.an.7.2	&	 				&	{\scriptsize Classification of SEDs}		& 	 			&				&	{\large X}		\\
SED.an.8		&				&					&	{\scriptsize Astrophysics and cosmology}	&			 	& 	{\large X}	 	&				\\
$\searrow$	& 	SED.an.8.1	& 	 				&	{\scriptsize Apply extinction laws}		&	 			& 	{\large X} 		&				\\
		$|$	& 	$\searrow$	&	SED.an.8.1.1	 	&	{\scriptsize User-provided ext. laws}		&	 			& 	{\large X} 		&				\\
		$|$	& 			$|$	& 	SED.an.8.1.2	 	&	{\scriptsize Libraries of ext. laws}		&	 			& 	{\large X}	 	&				\\
		$|$	&			$|$	& 	SED.an.8.1.3	 	&	{\scriptsize Tabular descript. of ext. laws}	&			 	& 			 	&	{\large X}		\\
		$|$	&			$|$	& 	SED.an.8.1.4	 	&	{\scriptsize Redshift-depend. ext. laws}	&	 			& 			 	&	{\large X}		\\		
		$|$	& 	SED.an.8.2	&	 				&	{\scriptsize Modify redshift of SED}		& 	 			&	{\large X}		&				\\
		$|$	& 	SED.an.8.3	&	 				&	{\scriptsize Estimate bolometric flux}		& 	 			&	{\large X}		&				\\
		$|$	& 	SED.an.8.4	&	 				&	{\scriptsize Bolometric lum. with distance}	& 	 			&	{\large X}		&				\\
SED.an.9		&				&					&	{\scriptsize Convolution}				&	 			& 	{\large X} 		&				\\
\end{tabular}
\label{table:prioritizationSEDan}
\end{table}

\begin{table}
\caption{Hierarchical break-down of the requirements and proposed prioritization of the deliverables for the ``SED visualization and interaction" section of the document.}
\centering
\begin{tabular}{r r c c c c c c}
\hline\hline
Task & Sub-task & Sub-sub-task & Description & Year 1 & Year 2 & Later... \\
\hline
SED.vis.1		&				&					&	{\scriptsize Plot the SED}				&	 {\large X}		& 			 	&				\\
$\searrow$	& 	SED.vis.1.1	& 	 				&	{\scriptsize Change coordinates of plot}	&	 {\large X}		& 			 	&				\\
		$|$	& 	SED.vis.1.2	&	 				&	{\scriptsize Replot spectral interv.}		& 	 {\large X}		&				&				\\
		$|$	& 	SED.vis.1.3	&	 				&	{\scriptsize Zoom in and out}			& 	 {\large X}		&				&				\\
		$|$	& 	SED.vis.1.4	& 	 				&	{\scriptsize Flag points and segments}	&	 			& 	{\large X}	 	&				\\
		$|$	& 	SED.vis.1.5	&	 				&	{\scriptsize Move points and segm.}		& 	 			&	{\large X}		&				\\
		$|$	& 	SED.vis.1.6	&	 				&	{\scriptsize Adjust data elements}		& 			 	&				&	{\large X}		\\
		$|$	& 	SED.vis.1.7 	&	 				&	{\scriptsize Inspect metadata}			& 	{\large X} 		&				&				\\
		$|$	& 	$\searrow$	&	SED.vis.1.7.1	 	&	{\scriptsize See m.d. with pointer}		&	{\large X} 		& 			 	&				\\
		$|$	& 			$|$	& 	SED.vis.1.7.2	 	&	{\scriptsize Save m.d. to file}			&			 	& 	{\large X}	 	&				\\
		$|$	&			$|$	& 	SED.vis.1.7.3	 	&	{\scriptsize Open links in m.d. in browser}	&			 	& 	{\large X} 		&				\\
		$|$	& 	SED.vis.1.8	&	 				&	{\scriptsize Export SED to VO-format}	& 			 	&	{\large X}		&				\\
		$|$	& 	SED.vis.1.9	&	 				&	{\scriptsize Plot mod. SED in new wind.} 	& 	{\large X}		&				&				\\
		$|$	& 	SED.vis.1.10	& 	 				&	{\scriptsize Interactive math on the SED}	&	 			& 			 	&	{\large X}		\\
		$|$	& 	SED.vis.1.11	& 	 				&	{\scriptsize Save plots in diff. formats}	&	{\large X} 		& 			 	&				\\
SED.vis.2		&				&					&	{\scriptsize Session report and CLI}		&	{\large X} 		&		 	 	&				\\
$\searrow$	& 	SED.vis.2.1	& 	 				&	{\scriptsize Complete session report}	&			 	& 	{\large X}	 	&				\\
		$|$	& 	SED.vis.2.2	&	 				&	{\scriptsize Simple session report}		& 	{\large X}		&				&				\\
		$|$	& 	SED.vis.2.3	&	 				&	{\scriptsize Workflow script for CLI}		& 			 	&				&	{\large X}		\\
SED.vis.3		&				&					&	{\scriptsize Plot of tabular data}			&	 			& 	{\large X}	 	&				\\
$\searrow$	& 	SED.vis.3.1	& 	 				&	{\scriptsize Scatter-plots}				&	 			& 	{\large X} 		&				\\
		$|$	& 	SED.vis.3.2	&	 				&	{\scriptsize Histograms}				& 	 			&	{\large X}		&				\\
		$|$	& 	SED.vis.3.3	&	 				&	{\scriptsize Plot of spatial modes/apert.}	& 	 			&				&	{\large X}		\\		
\end{tabular}
\label{table:prioritizationSEDvis}
\end{table}
\end{document}